\documentclass[journal]{IEEEtran}

\usepackage{hyperref}
\usepackage{array} % Table sizes
\usepackage{longtable}
\usepackage{booktabs} % pretty tables
\usepackage{makecell}
\usepackage{diagbox}
\usepackage{subfig}
\usepackage{amssymb}
\usepackage{graphicx}

\usepackage{amsfonts}

\usepackage{amsmath}
\usepackage{latexsym}
\usepackage{epsfig}
\usepackage{moreverb}
\usepackage{rotating}
\usepackage{enumerate}
\usepackage{graphics, graphicx,wrapfig}
\usepackage{fancybox}
\usepackage{picinpar,varioref,floatflt}
\usepackage{ae}
\usepackage{textcomp}
\usepackage{float}
\usepackage{lscape}
\floatstyle{plaintop}
\restylefloat{table}
\usepackage{caption}
\usepackage{multirow}

\usepackage{algpseudocode}
\usepackage{algorithm}
\usepackage{braket}

\begin{document}

\title{RL and Fingerprinting to Select Moving Target Defense Mechanisms for Zero-day Attacks in IoT}

\author{Alberto Huertas Celdr\'an$^{*1}$, Pedro Miguel S\'anchez S\'anchez$^{2}$, Jan von der Assen$^{1}$, Timo Schenk$^{1}$,\\ G\'er\^ome Bovet$^{3}$, Gregorio Mart\'inez P\'erez$^{2}$, and Burkhard Stiller$^{1}$

\thanks{$^{*}$Corresponding author.}

\thanks{$^{1}$Alberto Huertas Celdr\'an, Jan von der Assen, Timo Schenk, and Burkhard Stiller are with the Communication Systems Group (CSG) at the Department of Informatics (IfI), University of Zurich UZH, 8050 Zürich, Switzerland {\tt\small (e-mail: huertas@ifi.uzh.ch; vonderassen@ifi.uzh.ch; timo.schenk@uzh.ch;  stiller@ifi.uzh.ch}).}
\thanks{$^{2}$Pedro Miguel S\'anchez S\'anchez and Gregorio Mart\'inez P\'erez are with the Department of Information and Communications Engineering, University of Murcia, 30100 Murcia, Spain {\tt\small (pedromiguel.sanchez@um.es; gregorio@um.es)}.}%
\thanks{$^{3}$G\'{e}r\^{o}me Bovet is with the Cyber-Defence Campus within armasuisse Science \& Technology, 3602 Thun, Switzerland {\tt\small (gerome.bovet@armasuisse.ch)}.}}

\markboth{Journal of IEEE Transactions on Information Forensics and Security ,~Vol.~XX, No.~X, XX~202X}%
{Shell \MakeLowercase{\textit{et al.}}: Bare Demo of IEEEtran.cls for IEEE Journals}

\maketitle

\begin{abstract}
Cybercriminals are moving towards zero-day attacks affecting resource-constrained devices such as single-board computers (SBC). Assuming that perfect security is unrealistic, Moving Target Defense (MTD) is a promising approach to mitigate attacks by dynamically altering target attack surfaces. Still, selecting suitable MTD techniques for zero-day attacks is an open challenge. Reinforcement Learning (RL) could be an effective approach to optimize the MTD selection through trial and error, but the literature fails when i) evaluating the performance of RL and MTD solutions in real-world scenarios, ii) studying whether behavioral fingerprinting is suitable for representing SBC's states, and iii) calculating the consumption of resources in SBC. To improve these limitations, the work at hand proposes an online RL-based framework to learn the correct MTD mechanisms mitigating heterogeneous zero-day attacks in SBC. The framework considers behavioral fingerprinting to represent SBCs' states and RL to learn MTD techniques that mitigate each malicious state. It has been deployed on a real IoT crowdsensing scenario with a Raspberry Pi acting as a spectrum sensor. More in detail, the Raspberry Pi has been infected with different samples of command and control malware, rootkits, and ransomware to later select between four existing MTD techniques. A set of experiments demonstrated the suitability of the framework to learn proper MTD techniques mitigating all attacks (except a harmfulness rootkit) while consuming $<$1 MB of storage and utilizing $<$55\% CPU and $<$80\% RAM.
\end{abstract}


\IEEEpeerreviewmaketitle

\section{Introduction}
\label{sec:introduction}

% IoT growth and SBC % Cybersecurity concerns & need for novel and automatic solutions
\IEEEPARstart{T}{}he Internet of Things (IoT) has experienced explosive growth over recent years, and forecasts estimate that the number of connected devices will continue growing by billions annually~\cite{riad2020dynamic}. IoT devices permeate many areas of modern society, with applications spreading from smart cities to healthcare. In such a context, single-board computers (SBC), like Raspberry Pis, deserve special attention due to their applicability, flexibility, price, support, and peripherals availability. However, the connectivity and resource-constrained nature of SBCs, together with the heterogeneity of their application scenarios, have also accelerated the emergence of cyberattacks affecting these devices \cite{stellios2018survey}. Recent cybersecurity studies have shown how cyberattacks affecting resource-constrained devices are increasing every year~\cite{alsheikh2021state}. Analyzing the problem in more detail, it can be seen that cybercriminals are moving towards zero-day attacks executed by malware that combines heterogeneous malicious behaviors such as remote control, data leakage, encryption, mining, code execution hiding, and other hostile actions~\cite{parrend2018foundations}. This trend complicates the challenge of defending SBCs, since detection and mitigation mechanisms must be varied and powerful.

% Moving Target Defense and RL as a promising solution
Assuming that perfect security is most likely not achievable, a novel cyberdefense paradigm called Moving Target Defense (MTD) was introduced in 2009~\cite{cai2016moving}. MTD aims to thwart adversaries by proactively or reactively moving specific system parameters (such as IP addresses, file extensions, or libraries) to prevent and defend against attacks~\cite{navas2020mtd}. However, in reactive scenarios, selecting optimal MTD techniques to mitigate ongoing zero-day attacks launched by malware is challenging~\cite{cho2020toward} First, attacks must be detected. For that, unsupervised anomaly detection (AD) systems based on machine learning (ML) have proven their effectiveness \cite{celdran2022intelligent}. The problem is that AD systems do not distinguish between malicious behaviors, and attack families have heterogeneous impacts and therefore their mitigation require different MTD techniques. In this sense, the combination of device behavioral fingerprint (considering CPU, RAM, or file systems consumption) and Reinforcement Learning (RL) could be used for learning, in an online fashion, the best mitigation for each attack. More in detail, an RL-based agent could learn by trial and error which MTD mechanism mitigates each zero-day attack according to the device behavioral differences before and after the MTD is deployed.

% Challenges
Despite the potential of using RL and MTD to mitigate zero-day attacks, the literature presents some open challenges. First, there is an evident lack of work evaluating RL and MTD solutions in resource-constrained devices and real-world scenarios. Most often, agents are trained in simulated environments without evaluating and transferring the learned policies to real contexts for validation. Second, the literature has not studied whether device behavioral fingerprinting is a practical approach to be combined with RL and precisely represent SBCs' states. Despite ransomware and botnets (to mention two of them) affect the SBCs' behavior (CPU, RAM, network, or file system, among others) differently, related work has not studied whether the dimensional space and stability of fingerprints allow online learning for cyberattack mitigation. Last but not least, no results in terms of consumption of resources are available in the literature to judge the feasibility of combining fingerprinting, RL and MTD in real scenarios with resource-constrained devices. 

With the aim of improving the previous limitations, this work has the following contributions:

\begin{itemize}
    \item The design and implementation of an RL-based framework able to learn the correct MTD mechanisms, mitigating heterogeneous zero-day attacks affecting SBC. The framework uses device behavioral fingerprinting to represent states of SBC affected by malware and RL to learn the best MTD technique per malware attack by trial and error. The learning process is driven by the Deep Q-Learning algorithm, which uses a reward function based on the predictions of an unsupervised ML AD system. The framework is publicly available as an open source project in~\cite{owngit}.

    \item The deployment of the framework on a Raspberry Pi 3 Model B acting as a spectrum sensor of a real-world crowdsensing platform called ElectroSense~\cite{electrosense}. The Raspberry Pi has been infected with three command and control-based malware (C\&C), two rootkits, and one ransomware. Then, the following existing MTD techniques have been considered for mitigation: \textit{IP shuffling}, effective against C\&C attacks; \textit{Library sanitizing}, relevant for rootkits; and two MTD techniques dealing with ransomware called \textit{Ransomware trap} and \textit{File randomization}.   
    
    \item The evaluation of the framework in terms of agent learning performance and resource consumption when deployed on the Raspberry Pi. A set of experiments has demonstrated the suitability of the framework to learn proper MTD techniques mitigating all attacks (except a passive rootkit behaving harmfulness) while consuming $<$1 MB of storage and utilizing $<$55\% of CPU and $<$80\% of RAM.

\end{itemize}

The remainder of this article is organized as follows. Section~\ref{sec:related} reviews related work combining RL and MTDs to mitigate IoT malware. Then, Section~\ref{sec:solution} motivates the problem tackled by this work and the design and implementation details of the proposed RL-based framework. Section~\ref{sec:experiments} evaluates the framework performance in terms of learning performance and resource consumption. Finally, Section~\ref{sec:conclusions} draws conclusions and future steps.

\section{Related work}
\label{sec:related}

This section reviews related work combining RL and MTDs to mitigate cyberattacks in different device types. In line with this, \cite{cho2020toward} found that it has not yet been investigated how to deploy among multiple MTD techniques optimally. However, there are some works considering the application of RL to optimize a single MTD technique.

% Configuration Selection
In this context, \cite{dass2021reinforcement} presented an RL-based approach to generate a diverse and secure set of software configurations for general MTD. The authors formulated the MTD strategy as a single-player game using Monte Carlo Prediction and tested the system success in finding secure configurations. Another approach is DQ-MOTAG~\cite{chai2020dq}, an anti-DDoS system combining Deep RL (DRL) and proactive network address shuffling MTD to block bot-like behavior. The RL-based algorithm of DQ-MOTAG adaptively adjusts the shuffling periods of the MTD technique to reduce network resource consumption while maintaining defense performance. DQ-MOTAG was evaluated in a simulated environment of jikecloud servers. The authors of~\cite{soussi2021moving} used Deep RL (DRL) for MTD action selection based on continuous monitoring of the network state. DRL was deployed either in a single-agent setup with an MTD controller for proactive defense or in a multi-agent setup with a game theoretic model, including an attacker and a defender for reactive defense. This work does neither provide implementation nor evaluation details.

% Attacker-Defender Multi-Agent RL
Several works have modeled RL for MTD as an adaptive multi-agent process between an attacker and a defender using game theory. For example, \cite{eghtesad2020adversarial} proposed an approach formulated as a two-player general-sum game of an attacker and a defender competing to control a set of servers. Deep Q-Learning and the Double Oracle Algorithm were used to derive an optimal MTD policy. As in many game-theoretic settings, the results were evaluated in a simulated manner. Similarly, the authors of \cite{sengupta2020multi} followed a game-theoretic multi-agent RL approach. They leveraged Bayesian Stackelberg Markov Games (BSMGs) to model uncertainty over attacker types and MTD specifics. Then, the BSS-Q algorithm was used to learn optimal movement policies. The authors evaluated the system in a simulated Web application scenario where databases and the programming language were used as moving parameters. The authors of \cite{yoon2021desolater} also employed multi-agent DRL for proactive IP shuffling MTD against reconnaissance attacks in in-vehicle SDNs. Their system aimed to minimize security vulnerability while maximizing service availability by changing link bandwidth allocation and the frequency of IP shuffling. The evaluation was performed as a proof-of-concept in an in-vehicle SDN prototype with three agents executing different controlling tasks.

\tablename~\ref{tab:related-work} summarizes the most important aspects of the previous works and compares them with the contribution of this work. As can be seen, there is a lack of research utilizing RL and MTD in real-world scenarios with resource-constrained devices such as SBC. Most existing solutions consider multi-agent setups, relying on game-theoretic models that simulate attackers and defenders. Finally, to the best of our knowledge, the usage of RL to select the correct MTD strategy among multiple MTD techniques, as done in this work, has not yet been studied in the literature.

\begin{table*}[ht!]
\caption{Overview of Related Work (Op/Operation of MTD: P-Proactive, R-Reactive. Env/Environment: S-Simulation, R-Real World)}
\label{tab:related-work}
\resizebox{\textwidth}{!}{%
\begin{tabular}{llllllllll|}
  \textbf{Work} &
  \textbf{Device} &
  \textbf{Attacks} &
  \textbf{MTD} &
  \textbf{RL Approach} &
  \textbf{Op} &
  \textbf{Env} &
  \begin{tabular}[c]{@{}l@{}}\textbf{State Data (S) / Actions (A)}\end{tabular} &
  \textbf{Reward} \\
  \hline \hline

    \begin{tabular}[c]{@{}l@{}}\cite{dass2021reinforcement}\\2021\end{tabular} &
    Computer &
    Unspecific &
    Unspecific &
    Classic Monte Carlo &
    R/P &
    S &
    \begin{tabular}[c]{@{}l@{}}S: System parameters\\ A: Parameters change\end{tabular} &
    \begin{tabular}[c]{@{}l@{}}Dependent on parameter\\fitness score\end{tabular} \\  \hline

    \begin{tabular}[c]{@{}l@{}}\cite{chai2020dq}\\2020\end{tabular} &  Computer &
    DDoS &
    Network &
    \begin{tabular}[c]{@{}l@{}}DRL\end{tabular} &
    P &
    S &
    \begin{tabular}[c]{@{}l@{}}S: Network state vector\\ A: Adapt shuffling period\end{tabular} &
    \begin{tabular}[c]{@{}l@{}} Dependent on attacks,\\ connections, and mitigations\end{tabular} \\ \hline

    \begin{tabular}[c]{@{}l@{}}\cite{soussi2021moving}\\ 2021\end{tabular} &
    Computer &
    \begin{tabular}[c]{@{}l@{}} DDoS, MitM, \\ Spoofing \end{tabular} &
    Network &
    \begin{tabular}[c]{@{}l@{}} Single Agent (Proactive)\\ Multi-Agent (Reactive)\end{tabular} &
    P/R &
    None &
    None &
    None \\ \hline
  
    \begin{tabular}[c]{@{}l@{}}\cite{eghtesad2020adversarial}\\2020 \end{tabular} &
    Computer &
    \begin{tabular}[c]{@{}l@{}}Control\\ over servers\end{tabular} &
    Unspecific &
    \begin{tabular}[c]{@{}l@{}}Game Theory,\\Two-player general sum game,\\Multi-Agent RL, DQ-Learning\end{tabular} &
    R &
    S &
    \begin{tabular}[c]{@{}l@{}}S: Server role and status\\ A: Reconfigure and probe \end{tabular} &
    \begin{tabular}[c]{@{}l@{}} Dependent on servers\\confidentiality and availability\\ \end{tabular} \\ \hline

    \begin{tabular}[c]{@{}l@{}}\cite{sengupta2020multi}\\2020\end{tabular} &
    Computer &
    \begin{tabular}[c]{@{}l@{}}Web App \\ attacks\end{tabular} &
    Platform &
    \begin{tabular}[c]{@{}l@{}}Game Theory, Multi-Agent RL \\ Attacker-Defender, BSS-Q\end{tabular} &
    R &
    S &
    \begin{tabular}[c]{@{}l@{}}S: System configuration\\A: Configuration change \end{tabular} &
    \begin{tabular}[c]{@{}l@{}}Dependent on attack impact \end{tabular} \\ \hline

    \begin{tabular}[c]{@{}l@{}}\cite{yoon2021desolater}\\2021\end{tabular} &
    \begin{tabular}[c]{@{}l@{}} Vehicle \\(sensors,\\ actuators)\end{tabular} &
    Reconnaissance &
    Network &
    \begin{tabular}[c]{@{}l@{}}DRL, Multi-Agent recurrent \\ deterministic policy gradient\\ with Anomaly Detection\end{tabular} &
    P &
    S &
    \begin{tabular}[c]{@{}l@{}}S: Network statistics, shuffling\\overhead, and vulnerabilities\\ A: IP shuffling\end{tabular} &
    \begin{tabular}[c]{@{}l@{}}Dependent on bandwidth\\allocation efficiency, security,\\and packet loss\end{tabular} \\ \hline

    \begin{tabular}[c]{@{}l@{}}Ours\\2022\end{tabular} &
    \begin{tabular}[c]{@{}l@{}}SBC, \\Raspberry Pi \end{tabular}&
    \begin{tabular}[c]{@{}l@{}}Rootkits, C\&C,\\ Ransomware\end{tabular} &
    \begin{tabular}[c]{@{}l@{}}Data, \\Libraries,\\ Network\end{tabular} &
    DRL and Anomaly Detection &
    R &
    R &
    \begin{tabular}[c]{@{}l@{}}S: Device fingerprint\\ A: MTD Technique Selection\end{tabular} &
    \begin{tabular}[c]{@{}l@{}}Dependent on  effectiveness\\of selected MTD\end{tabular} 
    \\ \hline
\end{tabular}%
}
\end{table*}

\section{RL-based Framework for Mitigating Zero-day Attacks in SBC}
\label{sec:solution}

This section first motivates the challenge of mitigating zero-day attacks in resource-constrained devices. Then, it proposes an RL-based cybersecurity framework that learns the right MTD techniques to mitigate unseen or zero-day attacks. The framework source code is available in~\cite{owngit}. 

The task of selecting effective MTD techniques for different malware families is more or less complex depending on the attack novelty. Well-known malware can be detected and classified based on previous knowledge, such as signatures, rules, or ML-based classifiers. Once knowing the malware type, the next step is to select the MTD technique able to mitigate that particular malicious behavior. In this scenario, the defender must know the attack affecting the system and the MTD technique functionality. However, with zero-day attacks executed by novel and unseen malware, the selection task becomes much more complicated because the malware behavior is unknown. Therefore, selecting the proper mitigation technique cannot rely on previous knowledge, and it must be learned in real-time or online. This work focuses on improving the current status of this problem considering the following requirements. 

\begin{itemize}
    \item The defender has no prior knowledge about malware samples affecting the device (zero-day attacks).
    
    \item There is no predefined set of device behaviors or environmental states, since heterogeneous zero-day attacks might have different impacts on the device. 
    
    \item Malware behaviors are not obfuscated and their malicious actions are performed as in reality. 
    
    \item There is no unique MTD technique able to mitigate all attacks.
    
    \item The defender has at least one effective MTD technique per malware. However, there is no certainty that the device behavior returns to normality after the malware mitigation.
    
    \item The MTD selection mechanism must run on real and resource-constrained devices. Therefore, the consumption of CPU, RAM, and storage are important.
\end{itemize}

\begin{figure}[ht!]
\centering
\includegraphics[width=0.6\linewidth]{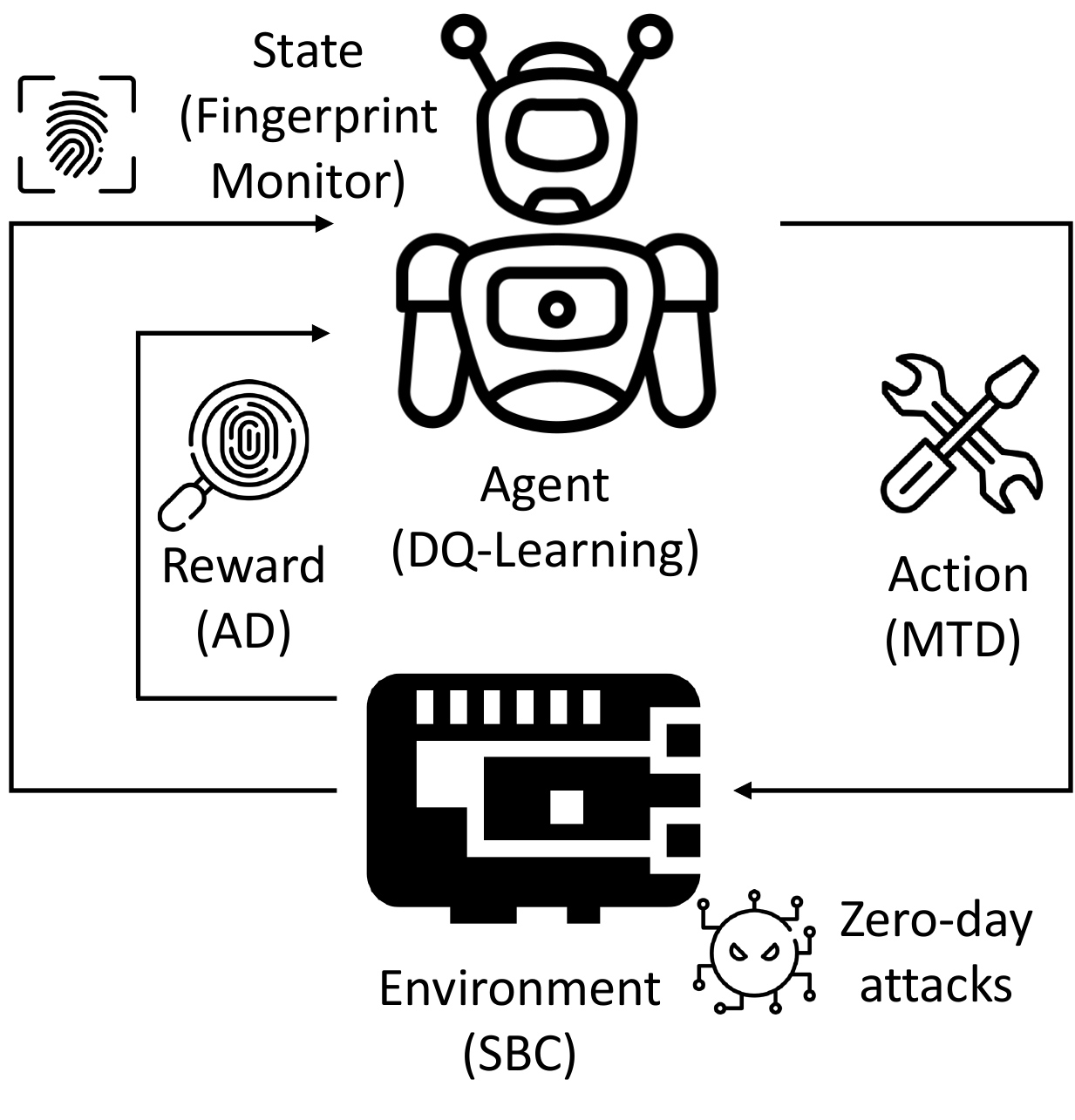}
\caption{RL-based Framework Overview}
\label{fig:framework}
\end{figure}

To fulfill the previous requirement, this work proposes a cybersecurity framework that uses RL to select proper MTD techniques depending on unknown attacks affecting SBC. \figurename~\ref{fig:framework} shows the main actors of the framework: the environment and the agent. In this work, the \textit{Environment} is an SBC affected by zero-day attacks, while the agent selects and deploys MTD techniques (known as actions) to mitigate the attacks. The learning process is done using Deep Q-Learning, the SBC state (interpreted by the agent as the device behavioral fingerprint), and the impact of the action on the environment (positive or negative reward calculated by an AD system). More details about both actors and the learning process are provided below.

\subsection{Environment}

%Raspberry as SBC
The environment considered in this work is an SBC. Particularly, a Raspberry Pi 3 model B with 1 GB RAM and Debian/Linux-based operating system (32-bit) running an ElectroSense Sensor~\cite{electrosense}. ElectroSense is a crowdsensing initiative that uses Raspberry Pis equipped with radio sensors to collect spectrum data worldwide and make this data available in real-time for analysis. In such a scenario, the Raspberry Pi is subject to three different families of malware: C\&C, rootkits, and ransomware. 

% Description of malware families and samples
C\&C malware establishes a communication channel between the compromised device and a server deciding harmful activity over the compromised one. Malicious functionality usually includes launching denial of service attacks, installing backdoors, spreading other malware, leaking sensitive data, or mining cryptocurrency. In particular, this work considers the following C\&C samples:

\begin{itemize}

    \item \textit{The tick}~\cite{thetickgit}. It allows for remote control of bots via a server using a remote shell and extracting files from victim devices.
    
    \item \textit{Backdoor\_jakoritar}~\cite{jakoritargit}. It provides a Python-based implementation for the client and server components, enabling data leakage and remote control.
    
    \item \textit{Backdoor\_dataleak}~\cite{owngit}. It consists of a shell script using the \textit{netcat} command to periodically leak sensitive information from files or commands such as \textit{ps aux}, \textit{ls /etc}, \textit{df -h}, or \textit{free}.
    
\end{itemize}

Regarding rootkits, user-level ones operate in user space, and their main goal is to preload malicious libraries specified by an attacker before legitimate ones can be loaded. In this sense, this work considers the following samples:

\begin{itemize}

    \item \textit{Beurk}~\cite{beurkgit}. It manipulates \emph{/etc/ld.so.preload} by appending malicious libraries. In this work, it is executed passively without harmful behavior.
    
    \item \textit{Bdvl}~\cite{bdvlgit}. It tampers with the location where the dynamic linker checks for preloading shared libraries (in the file \emph{/lib/arm-linux-gnueabihf/ld-2.24.so}). Instead of \emph{/etc/ld.so.preload}, it adds a custom path, which points to a malicious version.
    
\end{itemize}

The last family is ransomware, which encrypts files with sensitive or important data to force the victim to pay a ransom for deciphering data. This work uses:

\begin{itemize}

    \item \textit{Ransomware\_PoC}~\cite{ransomwaregit}. It is a crypto-ransomware that encrypts text, images, and video files containing sensitive information to ask later for a ransom.
    
\end{itemize}

\subsection{Agent}

The agent main goal is to learn effective MTD mechanisms mitigating each previous malware launched as zero-day attack (without prior agent knowledge). This learning task is performed online by interacting with the environment using RL. In RL-based agents, the following aspects are critical: state, action, and reward.

\subsubsection{State}

A state is the agent vision of the environment at a given time. Therefore, the complete set of states are all possible environmental representations over time. Precise states allow the agent to understand the environment and learn proper actions. In this work, since there are zero-day attacks, the events defining states must be generic enough to represent heterogeneous attacks. 

With that goal in mind, this framework uses device behavioral fingerprinting to represent environment states. In particular, software and kernel tracepoint events are considered because they cover an extensive range of promising events for attack detection and representation, as identified in previous works~\cite{celdran2022intelligent}. On the one hand, software events are based on low-level kernel counters, which comprise context switches and CPU migrations. On the other hand, kernel tracepoint events are static kernel-level instrumentation points that are hard-coded in relevant logical places of the kernel. The proposed framework uses the \textit{perf} Linux command to collect data belonging to these two data sources. \textit{perf} is a powerful and lightweight profiling tool that can be leveraged to monitor in-device event sources. 

\begin{figure*}[ht!]
\centering
\includegraphics[width=1\linewidth]{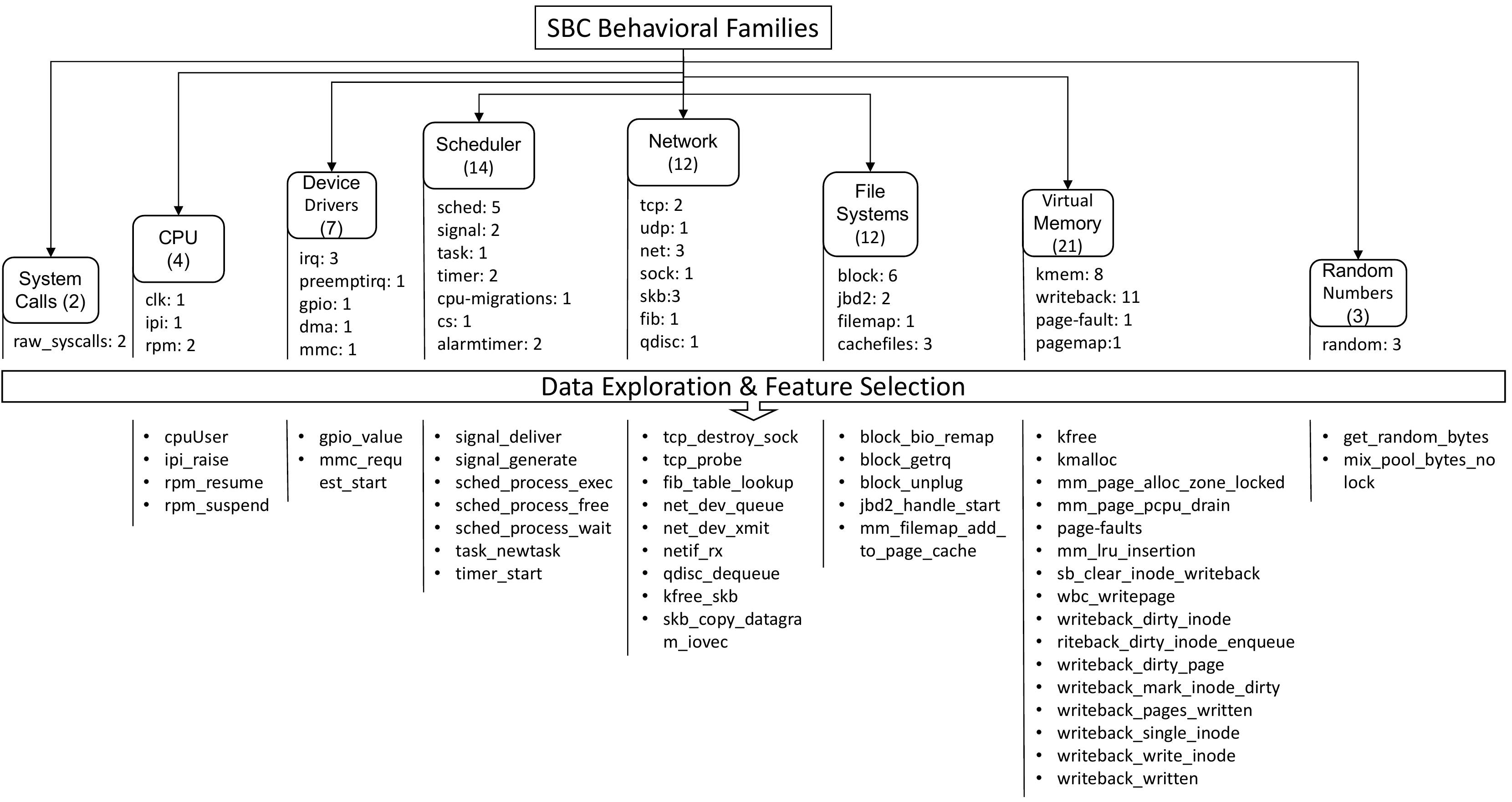}
\caption{Behavioral Families and Events Selected to Represent SBC States}
\label{fig:events}
\end{figure*}

Initially, 75 different \textit{perf} events belonging to \textit{system calls}, \textit{CPU}, \textit{device drivers}, \textit{scheduler}, \textit{network}, \textit{file system}, \textit{virtual memory}, and \textit{random numbers} families were monitored. The selection criterion was to cover as many different sources as possible to detect any small perturbation produced by heterogeneous zero-day attacks. At this point, it is important to mention that states or fingerprints must be precise enough and stable over time. Furthermore, the higher the number of events representing states, the higher the complexity of the state representation (feature dimensionality) and the longer the time the agent will take to converge during learning. Therefore, data exploration was performed to evaluate the suitability of the 75 features initially selected. All features were monitored in time windows of 5 s during 8 hours of Raspberry Pi normal behavior (sensor without being attacked). At this point, it is important to mention that previous work has demonstrated the suitability of the selected time window and monitoring duration \cite{sanchez2022studying}. Once the dataset was collected, data distributions of all features were analyzed, and features with constant or unstable values, or $>$90\% correlation, were removed. \figurename~\ref{fig:events} shows the families and number of events initially considered (8 families and 75 events) and the 46 events finally selected.

\subsubsection{Action}

Actions are how the agent interacts with the environment. In this work, actions correspond to the deployment of MTD techniques to mitigate zero-day attacks launched by C\&C, ransomware, and rootkits. The scope of this framework does not focus on proposing new MTD mechanisms but on a selection mechanism. Therefore, it considers the following MTD techniques proposed in~\cite{von2022lightweight}.

\begin{itemize}

    \item \textit{IP shuffling}. It is effective against C\&C attacks because it migrates the victim private IP address to a new one such that the control server cannot reach the victim anymore. 

    \item \textit{Ransomware trap}. It consists of expanding and collapsing the directory tree, creating dummy files that will be encrypted instead of good ones. In parallel, it identifies and kills the encryption process.  

    \item \textit{File randomization}. It is effective against malware dealing with files, like ransomware, and changes the file format extension to hide them from manipulation.

    \item \textit{Library sanitation}. It mitigates rootkits by shuffling between different sets of shared system libraries and cleans the links pointing to them.
    
\end{itemize}

The details about how the agent learns to select proper actions are provided after the Reward Section.

\subsubsection{Reward}

Positive and negative rewards let the agent learn if selected actions per state are good or bad. In this context, this work proposes the usage of an AD system based on unsupervised ML to automatize the rewards. More in detail, once a given attack affects the Raspberry Pi and a certain MTD technique is selected by the agent and deployed, the AD evaluates the new device behavior (state after the agent action, called afterstate). Then, if the AD prediction is normal, the deployed MTD mitigated the attack, and the reward is positive. In contrast, if the device behavior is abnormal, the selected MTD is ineffective against that attack, and the reward is negative.

To provide the previous functionality, first, an offline process trained an Autoencoder with the Raspberry Pi normal behavior. The Autoencoder has 46 neurons in the input and output layers (features representing the device fingerprint) and contains three hidden layers, with sizes 15, 7, and 15, respectively. All layers use Gaussian Error Linear Unit (GELU) as activation function. For training, the previously selected 46 events were monitored for eight days when the Raspberry Pi was not affected by any attack, and a dataset was created. Once having the dataset, the following tasks were performed: i) split the dataset into training and validation sets, ii) normalize feature values, and iii) eliminate outliers using the Z-score approach. After that, the Autoencoder was trained with 80\% of its dedicated normal data over 100 epochs with a batch size of 64 samples using Stochastic Gradient Descent with momentum as optimizer. Furthermore, the learning rate was $1e^{-4}$, and the momentum term was 0.9. The remaining 20\% of the samples were used to calculate the threshold as the mean predicted MSE reconstruction loss + 2.5 standard deviations and to find the previous configuration of hyperparameters.

Then, in real time, an online process evaluates each environment afterstate (state after deploying an MTD). For that, first, each of the previously mentioned malware samples is executed on the Raspberry Pi. Then, the agent is triggered (by the AD identifying the device state as abnormal) to select and deploy a given MTD technique. Once the selected MTD technique is deployed, and after giving the MTD two minutes to mitigate or not the attack, the AD evaluates again the current behavior (afterstate). If it is detected as normal, the selected MTD mitigates that attack, and the agent receives a positive reward (+1). If not, the reward is negative (-1).

\subsubsection{Learning Right Actions}

%Background about RL
The agent learns following a trial-and-error approach. In particular, when the agent takes one action for a given state, the environment feeds its new state (afterstate) and the reward back to the agent. The agent selects the following action based on this new information, and this loop is repeated. Sequences of the previous steps (state, action, afterstate and reward) are called episodes, which conclude when the agent cannot perform more actions. In this work, episodes are sequences of device fingerprints (states and afterstates), MTD mechanisms (actions), and rewards. An episode concludes when the effective MTD mechanism for the zero-day attack affecting the SBC is selected.

Mathematically speaking, the goal of the agent is to maximize the expected cumulative discounted future rewards: \begin{math} G_t=R_{t+1} + \gamma R_{t+2} + \gamma^2 R_{t+3} + ... = \sum_{k=0}^{\infty}R_{t+k+1} \end{math}. Where, \begin{math}R_{t}\end{math} denotes the reward at time step t, and \begin{math}\gamma\end{math} corresponds to a discount factor. To maximize this expected return \begin{math}G_t\end{math}, the agent needs to learn a policy. A policy is generally defined as a mapping from states to probabilities of selecting each possible action: \begin{math}\pi_t(a|s)\, \forall s \in States, a \in Actions \end{math}. During learning, through experience, these probabilities are shifted towards actions that lead to higher cumulative rewards. Experience in the form of a sequence of observed rewards can be captured by a so-called value function. A value function maps a state to a value, which is the estimated expected reward in that state. In general, RL-based methods aim to find the optimal policy by iteratively and alternatingly estimating value functions and improving a current policy. On the one hand, estimating the value function consists of predicting the state/action-state values, taking the current policy as fixed. How the values are predicted heavily depends on the concrete RL method applied. On the other hand, policy improvement is achieved by making the policy greedy for the current value function. 

In this work, the state space is large (46 features taking continuous values). Therefore, it is necessary to approximate the action-value function instead of learning it exactly and a purely computational method like Dynamic Programming~\cite{bellman2015applied} can be excluded. Instead, Deep RL is a promising approach, as it approximates the action-value function via a deep neural network. During the neural network training, there are different alternatives to decide when its parameters are updated. In Monte Carlo methods, rewards are only available at the end of episodes~\cite{mohamed2020monte}. It means that the neural network could only be updated at the end of an episode, increasing the learning time and making the training more unstable. Temporal Difference Learning~\cite{pong2018temporal} makes the training process faster since the action-value function can be updated after every action. Therefore, Deep Q-Learning is an adequate choice in this work, as it utilizes Temporal Difference and accounts for large state spaces. Further, it allows the agent to learn based on randomly sampled transitions from a replay memory, which ensures that frequently occurring sequences of attacks and actions are decorrelated. 

Algorithm \ref{alg:dq-learning} presents the variant of Deep Q-Learning used by the proposed framework to train the agent. As can be observed from the nested loop, learning happens over a number of $M$ episodes and $T$ time steps per episode. First, there is an action choice given a state $s_t$. This choice is based on an exploitation-exploration trade-off (lines 7-8). Next, an action is executed, and a tuple (with the state, action, reward, and next state) is stored in $D$, the replay memory (lines 9-10). Then, a batch is sampled from memory (line 11), and targets are calculated using a separate target network based on the temporal difference update. One particularity in Deep Q-Learning is the usage of two neural networks, one for the target calculation $Q^T$ (target network) and another for predicting the current action $Q^O$ (online network). The reason is to obtain more target stability and robustness during training. Then, as the core of the learning procedure, the online network is updated based on targets $y$ derived from a random sample of all transitions stored in $D$ (lines 11-14). In this sense, learning does not necessarily happen based on presently observed state-action pairs but on random samples replayed from memory. This is crucial for decorrelating sequences of state-action pairs that often occur in reality and avoiding updating the network in an unstable manner, making convergence difficult. The next key feature of the algorithm is the temporal difference target (line 13). Structurally, $y_j$ exactly corresponds to the update in classic, tabular Q-learning. The target q-value is calculated based on the current reward $r_j$ and the maximum-valued action. The maximum-valued action is further weighted by the so-called discount factor $\gamma$, which determines the importance of future rewards. Then a gradient descent step is performed over the online network. Finally, after a certain number of update steps (line 17-19, $update\_freq$), the weights of the online network $Q^O$ are copied over to the target network $Q^T$, and the next action is taken, or the next episode is started.

\begin{algorithm}[htpb]
\scriptsize
\caption{Deep Q-Learning with Experience Replay}\label{alg:dq-learning}
\begin{algorithmic}[1]

\State {Initialize replay memory $D$ to capacity N}
\State {Initialize online and target action-value functions $Q^O$ and $Q^T$ with random weights}
\State{Initialize exploration factor $\epsilon$ close to 1}
\For{episode = 1, M (max number of episodes)}
  \State {Initialize $s_t$}
  \For{t = 1, T (max timesteps within an episode)}
      \State{With probability $\epsilon$ select a random action $a_t$}
      \State{Otherwise select $a_t$ = $max_a(Q^O(s_t, a; \theta))$}
      \State{Execute action $a_t$ and observe reward $r_t$ and state $s_{t+1}$}
      \State{Store transition $(s_t, a_t, r_t, s_{t+1})$ in $D$}
      \State{Sample random batch of transitions $(s_t, a_t, r_t, s_{t+1})$ from $D$}
      \State{Calculate targets:}
      \State{\[y_j = \Set{\begin{array}{l}
        r_j \hfill \text{for terminal } s_{t+1} \\
        r_j + \gamma max_{a^{'}}(Q^T(s_{t+1}, a^{'}; \theta)) \hfill \; \text{for non-terminal} s_{t+1} \end{array}}\]}
      \State{Batch gradient descent step using $(y_j - Q^O(s_t, a_j; \theta))^2$}      
      \State{$s_{t}$ $\gets$ $s_{t+1}$}
      \State{Perform $\epsilon$-decay to minimize exploration over time}
      
      \If{$tot\_steps$ $\mod$ $update\_freq$ $==$ $0$} 
        \State{$Q^T$ $\gets$ $Q^O$, update target net}
      \EndIf
      
  \EndFor
\EndFor

\end{algorithmic}
\end{algorithm}

Regarding the agent hyperparameters, the input layer of the DQ-Network has 46 nodes, conforming the number of features representing states. The output layer has four nodes, corresponding to the available MTD techniques. Two hidden layers with 60 and 30 nodes complete the network. The replay memory is configured as a ring buffer with 500 transitions at maximum and is initialized with 100 sample transitions before the agent starts to learn. The batch size for the gradient descent step in the RL loop is set to 100 transitions. ADAM is the selected optimizer, $1e^{-4}$ is chosen as the learning rate, and the reward discount factor $\gamma$ is 0.1. Choosing $\gamma$ close to 0 ensures that immediate rewards are weighted much more than future rewards, which is desirable for correct MTD selection and speeds up the training process. $\epsilon$ as the exploration parameter starts at probability 1.0 and decays with every learning update by $1e^{-4}$ until it reaches a minimum level of 0.01 (ensuring that the agent continues to explore indefinitely). Finally, the frequency of replacing the target network $Q^T$ with the online network $Q^O$ is set to every 100 learning update steps. The previous configuration was chosen after an hyperparameter search.

\section{Experiments}
\label{sec:experiments}

This section presents a set of experiments to evaluate the performance of the RL-based framework on the Raspberry Pi. The metrics calculated in the experiments are: the agent selection performance, the AD performance, and the amount of disk, RAM and CPU of the Raspberry Pi consumed by the framework.

\begin{figure}[ht!]
\centering
\includegraphics[width=1\linewidth]{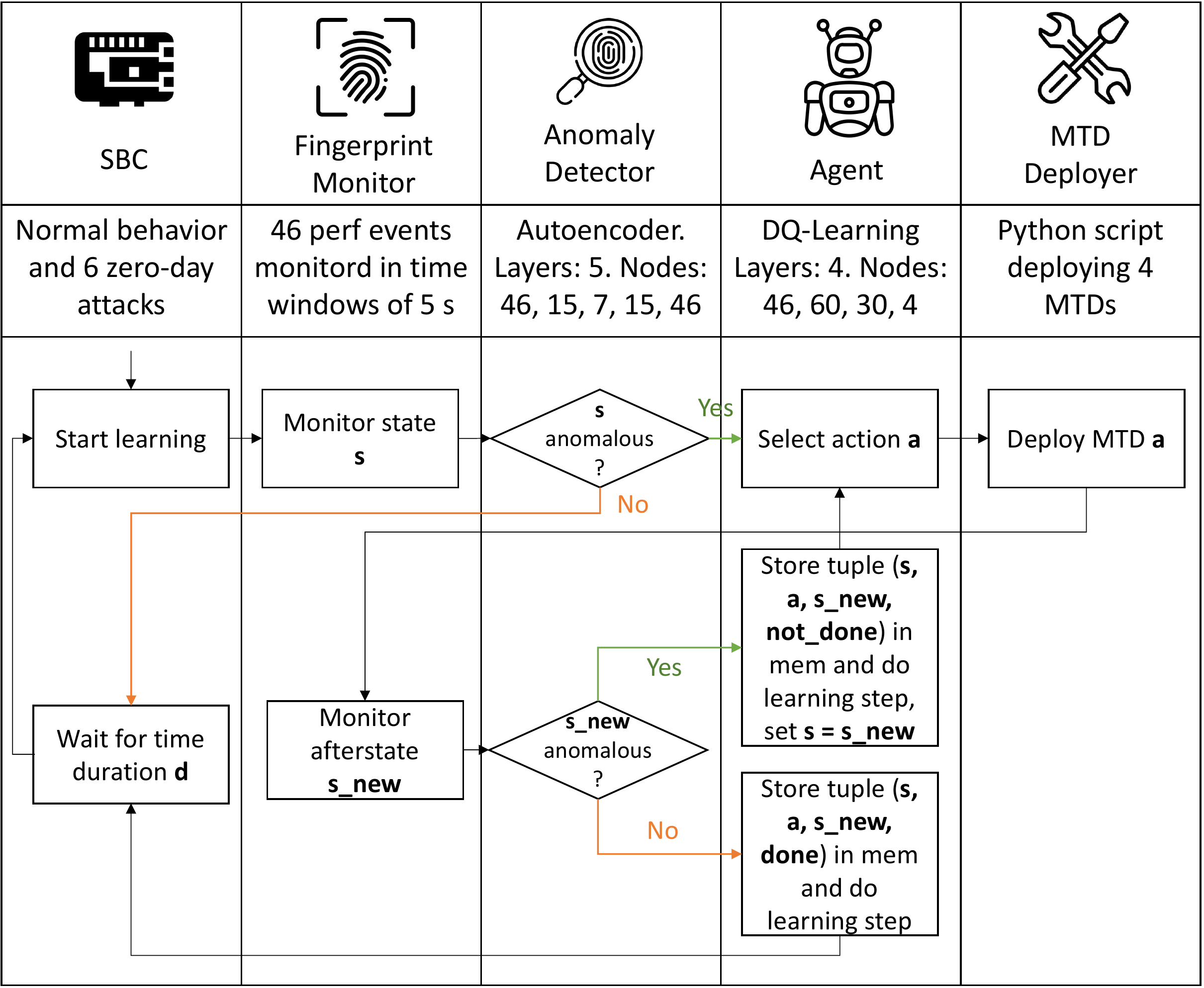}
\caption{Online Agent Learning Process Life Cycle}
\label{fig:lifecycle}
\end{figure}

\figurename~\ref{fig:lifecycle} displays the life cycle used to perform the experiments. When the learning process starts, the monitoring component first senses the Raspberry Pi behavior (\textit{state s}). Then, the AD component (already trained with the Raspberry Pi normal behavior) decides whether the current behavior is abnormal (or not). If so, the agent is triggered and selects one particular MTD technique (\textit{action a}), which is deployed on the Raspberry Pi. Then, after giving the MTD technique some time to perform its action (two minutes, due to the MTD implementation), the fingerprint monitor senses another fingerprint of the Raspberry Pi (\textit{afterstate s\_new}), which is evaluated by the AD. If the afterstate is abnormal, the agent receives a negative reward (-1), performs the DQ-learning update, and selects a new action. These steps are repeated while wrong actions are chosen. Finally, when the afterstate is normal (the attack is mitigated), the agent receives a positive reward (+1), updates its action-value network, and concludes the episode. A new episode starts after waiting until the device is infected by another attack.

\begin{figure}[ht!]
\centering
\includegraphics[width=1\linewidth]{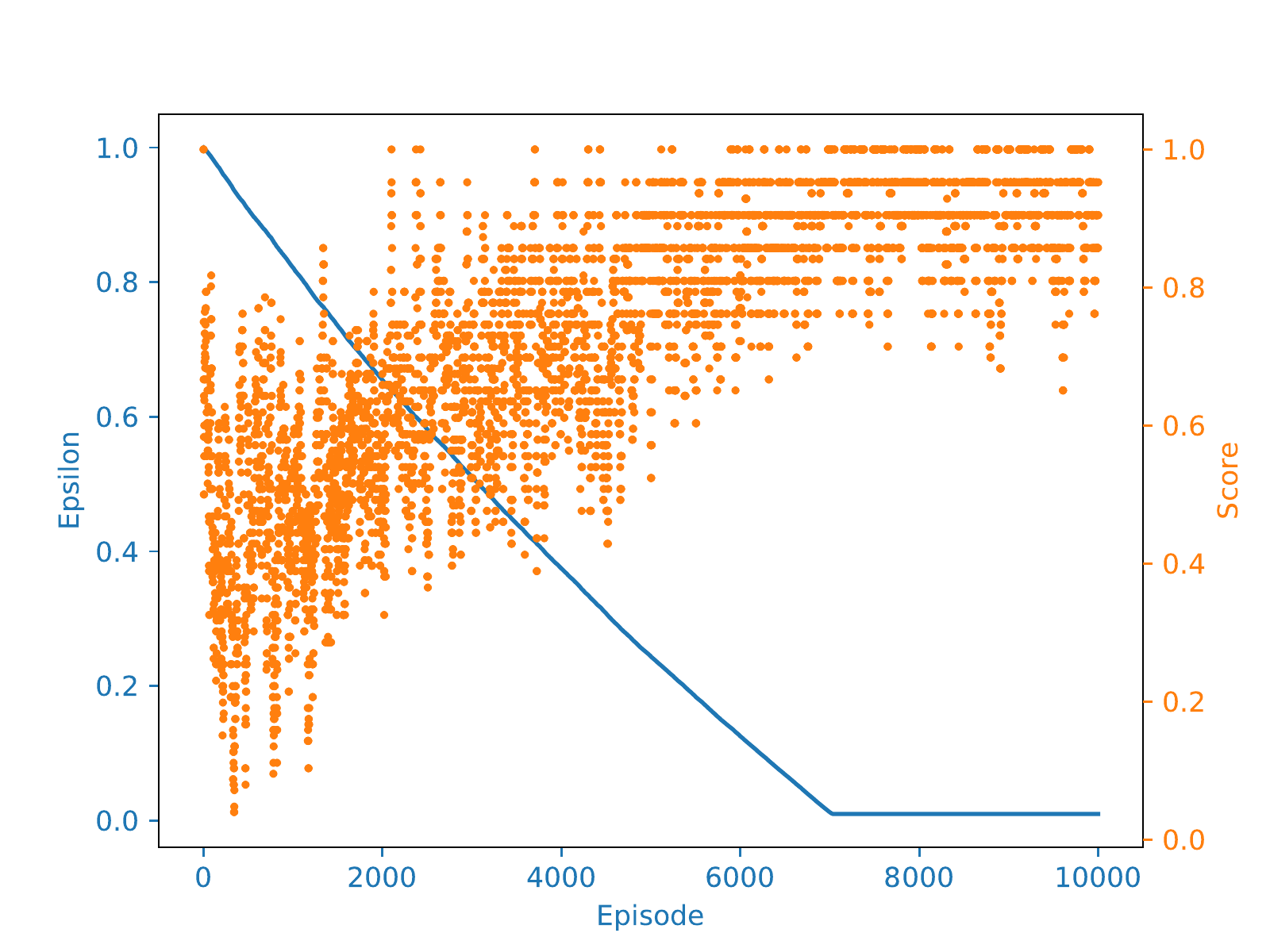}
\caption{Learning Over Episodes and Epsilon Decay}
\label{fig:learning}
\end{figure}

%Agent learning performance
To calculate the agent learning performance, before starting each episode, the execution begins with an 80\% probability of normal behavior and 20\% random attack. Then, the AD (Autoencoder) makes the normal/abnormal decisions based on a single SBC behavioral sample. \figureautorefname~\ref{fig:learning} shows the convergence of the agent towards the maximum episode reward for a total of 10000 episodes.

% explanation of variation
As can be seen, there is some variation in the learning process, but the tendency to approximate an optimal policy is clear and stable at $\approx$7000 episodes. This variation is due to two main challenges. First, the prediction of a correct MTD from a state. Secondly, the selection of the right MTD from an afterstate when the agent failed on states.

%Agent selecton performance
In order to analyze these two aspects, after finishing the training process, the accuracy of the agent greedy action choices is evaluated on a different execution. In this new execution, all states and afterstate are used to evaluate the agent greedy action selection according to its $Q^O$ online DQ-Network. \tablename~\ref{tab:agent_perf_state} presents the results on state and afterstate. Note that no meaningful accuracy can be evaluated for all behaviors considered normal. So, they have not been included in the table.

% agent accuracy
\begin{table}[htpb]
    \centering
    \scriptsize
    \caption{Agent Accuracy for States and Afterstates}
    \label{tab:agent_perf_state}
    \begin{tabular}{llll}
    
         \textbf{Behavior} & \textbf{Accuracy} & \textbf{Target Action} \\
        \hline \hline
         Ransomware\_PoC & 96.58\% & \begin{tabular}[c]{@{}l@{}} Ransomware trap,\\ File randomization\end{tabular} \\ \hline
         Ransomware\_PoC + IPS  & 100.00\% & \begin{tabular}[c]{@{}l@{}} Ransomware trap,\\ File randomization\end{tabular} \\ \hline
         Ransomware\_PoC + LS & 96.67\% & \begin{tabular}[c]{@{}l@{}} Ransomware trap,\\ File randomization\end{tabular}\\ \hline
         
         Bdvl & 93.15\% & Library sanitation  \\ \hline
         Bdvl + RT  & 94.93\% &  Library sanitation  \\ \hline
         Bdvl + FR & 91.87\% & Library sanitation  \\ \hline
         Bdvl + IPS  & 91.47\% & Library sanitation \\ \hline
         
         Beurk & 15.31\% & Library sanitation\\ \hline
         Beurk + RT  & 16.58\% & Library sanitation \\ \hline
         Beurk + FR & 15.40\% & Library sanitation \\ \hline
         Beurk + IPS  & 15.52\% & Library sanitation \\ \hline
         
         The tick & 63.92\% & IP shuffling \\ \hline
         The\_tick + RT  & 64.27\% & IP shuffling \\ \hline
         The\_tick  + FR & 63.37\% & IP shuffling \\ \hline
         The\_tick  + LS & 60.10\% & IP shuffling \\ \hline
         
         Backdoor\_jakoritar & 77.10\% & IP shuffling \\ \hline
         Backdoor\_jakoritar + RT  & 70.07\% & IP shuffling \\ \hline
         Backdoor\_jakoritar + FR & 73.25\% & IP shuffling \\\hline
         Backdoor\_jakoritar + LS & 60.48\% & IP shuffling \\\hline
        
         Backdoor\_dataleak & 100.00\% & IP shuffling \\ \hline
         Backdoor\_dataleak + RT & 100.00\% & IP shuffling \\ \hline
         Backdoor\_dataleak + FR & 99.76\% & IP shuffling\\ \hline
         Backdoor\_dataleak + LS  & 100.00\% & IP shuffling \\ \hline
    \end{tabular}
\end{table}

% agent performance
Looking at the states, the agent fails to learn the correct MTD for beurk ($\approx$15\% accuracy). It is important to mention that beurk is executed passively without performing any harmful action on the Raspberry Pi. In addition, backdoor\_jakoritar and the tick are mitigated with $\approx$77\% and $\approx$64\% accuracy, and all other attacks are correctly mapped to the right MTD technique (93-100\% accuracy). The agent performance on afterstates is very close to the states. More in detail, after deploying incorrect MTDs, the correct MTD technique is selected with 60\%-70\% accuracy for the tick and the backdoor, and 92\%-100\% for all other afterstates combinations, except those with beurk ($\approx$15\%).

In order to understand better the previous results, particularly the poor results obtained with beurk, the next experiment evaluates the AD performance on states and afterstates. The objective is to determine if erroneous rewards provided by the AD influence wrong agent actions or if the agent cannot distinguish between normal and zero-day attack states (or both). \tablename~\ref{tab:autoencoder_performance} shows the performance of the Autoencoder (AD) for all states and afterstate. The first column shows the Raspberry Pi behavior (state) and the deployed MTD technique (afterstate). The second and third columns show the AD accuracy when detecting that behavior as normal or abnormal. All afterstates with correct MTD techniques for a given attack should be recognized as normal, while all attacks with incorrect MTDs should be abnormal. These results are obtained evaluating $\approx$2000 samples per behavior on the Autoencoder explained in Section~\ref{fig:framework}.

\begin{table}[ht!]
    \centering
    \scriptsize
    \begin{tabular}{lll}
         \textbf{Behavior}  & \textbf{Accuracy} & \textbf{Target State}    \\
         \hline \hline
         Normal & 94.99\% & Normal \\ \hline
         %Normal + Ransomware trap & 95.76\% (n)  \\
         %Normal + File randomization & 94.29\% (n)  \\
         %Normal + Library sanitation & 93.79\% (n)   \\
         %Normal  + IP shuffling & 94.79\% (n) \\
         
         Ransomware\_PoC  & 100.00\% & Abnormal \\\hline
         Ransomware\_PoC + Ransomware trap & 93.33\% & Normal  \\\hline
         Ransomware\_PoC + File randomization & 94.21\% & Abnormal  \\\hline
         Ransomware\_PoC + IP shuffling & 100.00\% & Abnormal \\\hline
         Ransomware\_PoC + Library sanitation & 100.00\% & Abnormal  \\\hline
    
         Bdvl  & 100.00\% & Abnormal\\\hline
         Bdvl + Ransomware trap & 100.00\% & Abnormal\ \\\hline
         Bdvl + File randomization & 100.00\% & Abnormal\ \\\hline
         Bdvl + IP shuffling & 100.00\%  & Abnormal\ \\\hline
         Bdvl + Library sanitation & 88.92\% & Normal \\\hline

         Beurk & 5.09\% & Abnormal\ \\\hline
         Beurk + Ransomware trap & 5.68\% & Abnormal\ \\\hline
         Beurk + File randomization & 6.45\%  & Abnormal\ \\\hline
         Beurk + IP shuffling & 6.61\% & Abnormal\ \\\hline
         Beurk + Library sanitation & 92.91\% & Normal  \\\hline
    
         The\_tick & 6.50\% & Abnormal\ \\\hline
         The\_tick + Ransomware trap & 6.06\% & Abnormal\  \\\hline
         The\_tick + File randomization & 6.72\% & Abnormal\ \\\hline
         The\_tick + IP shuffling & 87.87\% & Normal \\\hline
         The\_tick + Library sanitation & 5.09\% & Abnormal\ \\\hline
    
         Backdoor\_jakoritar & 6.81\% & Abnormal\ \\\hline
         Backdoor\_jakoritar + Ransomware trap & 5.05\% & Abnormal\ \\\hline
         Backdoor\_jakoritar + File randomization & 12.06\% & Abnormal\ \\\hline
         Backdoor\_jakoritar + IP shuffling  & 91.57\% & Normal \\\hline
         Backdoor\_jakoritar + Library sanitation  & 6.63\% & Abnormal\ \\\hline
    
         Backdoor\_data\_leak & 100.00\% & Abnormal\ \\\hline
         Backdoor\_dataleak + Ransomware trap & 100.00\% & Abnormal\ \\\hline
         Backdoor\_dataleak + File randomization & 100.00\% & Abnormal\ \\\hline
         Backdoor\_dataleak + IP shuffling  & 88.35\% & Normal \\\hline
         Backdoor\_dataleak + Library sanitation & 100.00\% & Abnormal \\ \hline
    \end{tabular}
    \caption{AD Accuracy for States and Afterstates}
    \label{tab:autoencoder_performance}
\end{table}

Looking at the accuracy on states, beurk, backdoor\_jakoritar and the tick are recognized poorly with only 5\%-7\%. This is due to their proximity to the Raspberry Pi normal behavior. The other attacks (ransomware\_PoC, bdvl, and backdoor\_dataleak) are detected with perfect accuracy, and normal behavior is correctly recognized in $\approx$95\% of the cases. Regarding the accuracy on afterstates, two main aspects can be observed. First, the accuracy is close to the one achieved on states. If an incorrect MTD technique is deployed, the afterstate is recognized poorly in case of beurk, backdoor\_jakoritar and the tick (5\%-12\%), and with high accuracy for the remaining attacks. The second aspect is that in case of deploying correct MTD techniques, the behavior is correctly recognized for all behaviors (>87\% accuracy). This is a desirable result for the agent training because ensures that beurk, backdoor\_jakoritar and the tick are correctly mitigated at some point, even though there might be incorrect entries in the network replay memory.

According to the previous results, it can be concluded that the agent learning convergence seems relatively robust, despite having attacks (like beurk) that are not detected well. The behavior of beurk is similar to the normal one, which provokes incorrect rewards and complicates the agent differentiating it from normal behavior. In fact, beurk is a passive malware that does not perform malicious actions, so from a behavioral perspective, it is normal. This claim is supported by another experiment (not included by lack of room), where perfect rewards are provided for beurk (in a supervised manner), and the agent accuracy is 67\% (while for the rest of malware is almost 100\%). Dealing with the tick and backdoor\_jakoritar, despite the AD not providing excellent performance, since their behaviors are not as close as beurk to normal, the agent learns their effective MTDs after many episodes. The AD clearly detects the rest of the malware, and the agent has no problem learning the proper MTD for them.

%The last experiment deals with the consumption of storage, CPU, and RAM of the framework when it is deployed on the Raspberry Pi. 

%Accordingly, \tableautorefname~\ref{tab:tab-proc-times} displays processing times monitored by setting timestamps at dedicated locations in the agent controller code for each functional component.

%\begin{table}[h!]
%  \begin{center}
%    \caption{Processing Times}
%    \label{tab:tab-proc-times}
%    \begin{tabular}{ll}
%        \hline
%        \textbf{Framework Component}  & \textbf{Time} \\
%        \hline
%        Fingerprint Monitor\\decision/afterstate & 100.08s/100.07s \\ \hline
%        AD\\decision/afterstate & 0.025s/0.02s \\ \hline
%        Agent Action Choice\\$\epsilon$-greedy/DQN & 0.014s\\ \hline
%        MTD Execution\\agent action & ms-mins \\ \hline
%        Agent update\\DQN SGD on replay memory & 0.05s \\ 
%        \hline
%    \end{tabular}
%  \end{center}
%\end{table}

%\tableautorefname~\ref{tab:tab-proc-times} clearly shows that no stage of online, RL-based MTD is notably hampered by excessive processing times. This means that a pretrained agent as from the third simulation will be an effective and efficient measure against the considered families of malware.

\begin{figure}[ht!]
\centering
\includegraphics[width=1\linewidth]{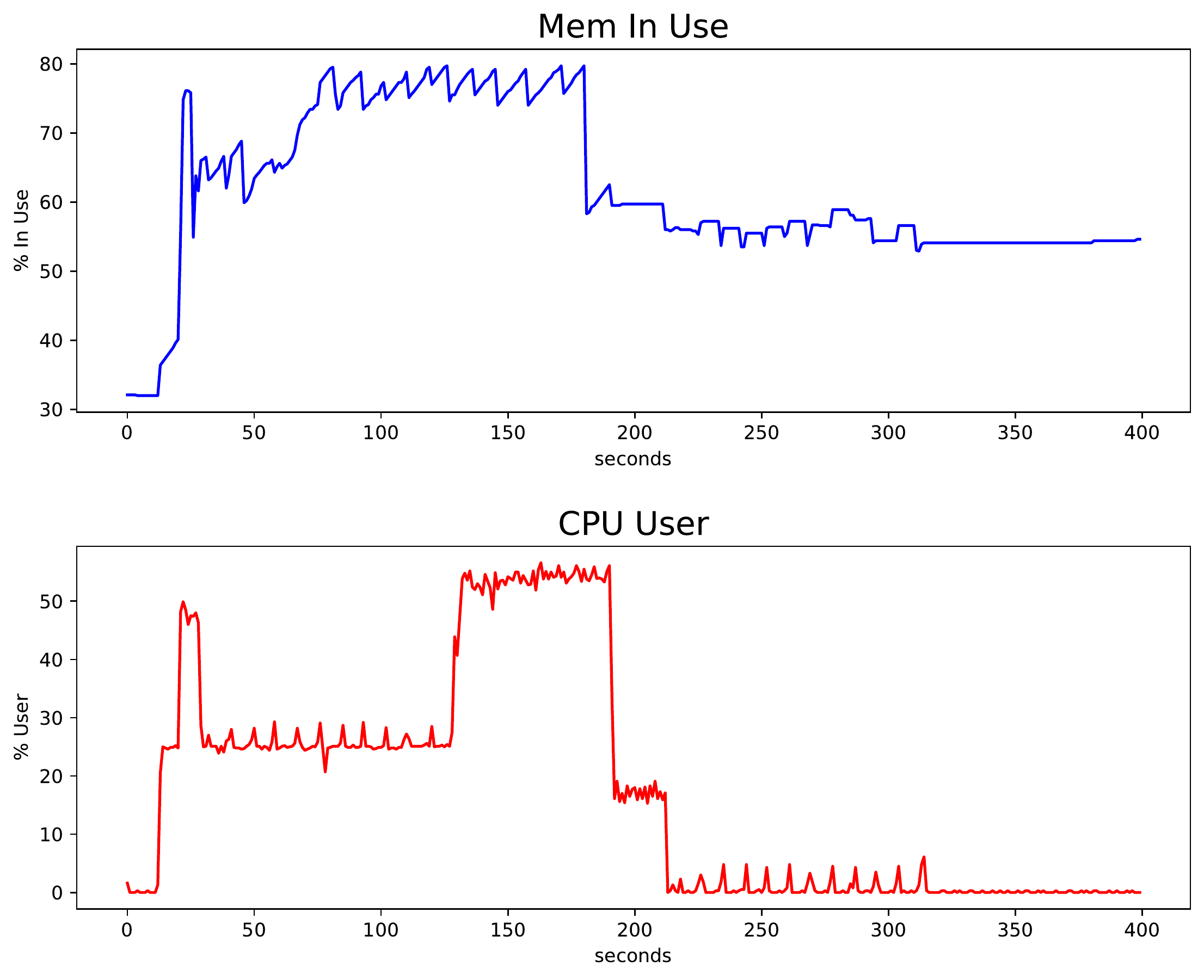}
\caption{RAM and CPU Used by the Framework for Ransomware\_PoC and Ransomware Trap MTD}
\label{fig:cpu_ram}
\end{figure}

The last experiment deals with the amount of disk, CPU, and RAM consumed by the framework when it is deployed on the Raspberry Pi. In terms of disk consumption, all MTD techniques and the agent require 964 KB. In addition, the Autoencoder needs 8 KB, and the datasets resulting from monitoring states and afterstates take 8 KB each. In order to assess framework CPU and RAM requirements, both are measured along the execution of the \textit{Ranssomware\_PoC} and the \textit{Directory trap} MTD (the most resource-consuming configuration). \figureautorefname~\ref{fig:cpu_ram} displays RAM and CPU User usage for this example.

Looking at RAM memory over time, it is possible to see the different stages of the agent execution. First, $\approx$30\% of the memory is in use as neither malware, nor agent code is running. Then, there is an increase to $\approx$70\% when the ransomware is launched against the RP. The maximum is reached at $\approx$80\% when the agent is actively learning. At about 120 s in the timeline, the Ransomware trap MTD is launched. Then, after mitigation, at $\approx$200 s, the level of memory in use stabilizes at $\approx$55\%. The percentage of CPU dedicated to user-level operations reaches its maximum at $\approx$55\%, after 140 s which is in line with the MTD execution. The maximum of CPU system usage is reached during the same time window with $\approx$25\%.

In conclusion, storage is certainly not a limiting factor of the RL-based framework when deployed on the Raspberry Pi. In terms of CPU and RAM, for the most resource-consuming attack and MTD mechanism that the agent faced, the Raspberry Pi does not seem to have any troubles executing all the agent processes in parallel. Thus, from a resource requirements perspective, the agent is fully functional without delays along its execution. Overall, it appears that from a resource consumption perspective, the orchestration functions of the agent are negligible compared to the actions taken by MTD techniques. As such, the agent presented here is able to optimize the execution of MTD techniques with minimal overhead.

\section{Conclusions}
\label{sec:conclusions}

This work explores the usage of RL and behavioral fingerprinting for selecting MTD techniques mitigating zero-day attacks in SBC. In particular, this work proposes an RL-based framework that considers 46 behavioral events to represent SBCs' states and DQ-Learning to learn effective MTD techniques to mitigate heterogeneous zero-day attacks. In addition, an AD system based on unsupervised ML provides positive or negative rewards to the MTD selected by the framework agent. The framework has been validated in a real scenario, covering an important gap in the literature, composed of a Raspberry Pi acting as a spectrum sensor of a crowdsensing platform called ElectroSense. The Raspberry Pi has been infected with six heterogeneous malware (C\&C, rootkits, and ransomware) in a zero-day manner. From the MTD perspective, four existing techniques have been considered to mitigate the previous attacks. A pool of experiments has evaluated the framework in terms of i) learning performance, where effective MTD techniques have been selected for all attacks except one running passively and without performing malicious actions, and ii) CPU, RAM, and storage consumption, needing $<$1 MB of storage and utilizing $<$55\% CPU and $<$80\% RAM of a Raspberry Pi 3. In conclusion, this work has improved the literature challenges by proving that RL and behavioral fingerprinting can be used in real scenarios with resource-constrained devices to select MTD mechanisms effective against zero-day attacks.

In future work, a larger amount of attacks and MTD techniques will be considered in the current RL-based MTD system. In addition, it is planned to analyze how long a pre-trained agent takes to learn new, unseen attacks. Another future work is to monitor other behavioral dimensions of SBC to create more precise environmental states and improve the detection of malware acting passively, as Beurk does in this work.

\section*{Acknowledgment}

This work has been partially supported by \textit{(a)} the Swiss Federal Office for Defense Procurement (armasuisse) with the CyberTracer and RESERVE projects (CYD-C-2020003) and \textit{(b)} the University of Zürich UZH.

\bibliographystyle{IEEEtran}
\bibliography{references}

\begin{IEEEbiography}[{\includegraphics[width=1in,clip]{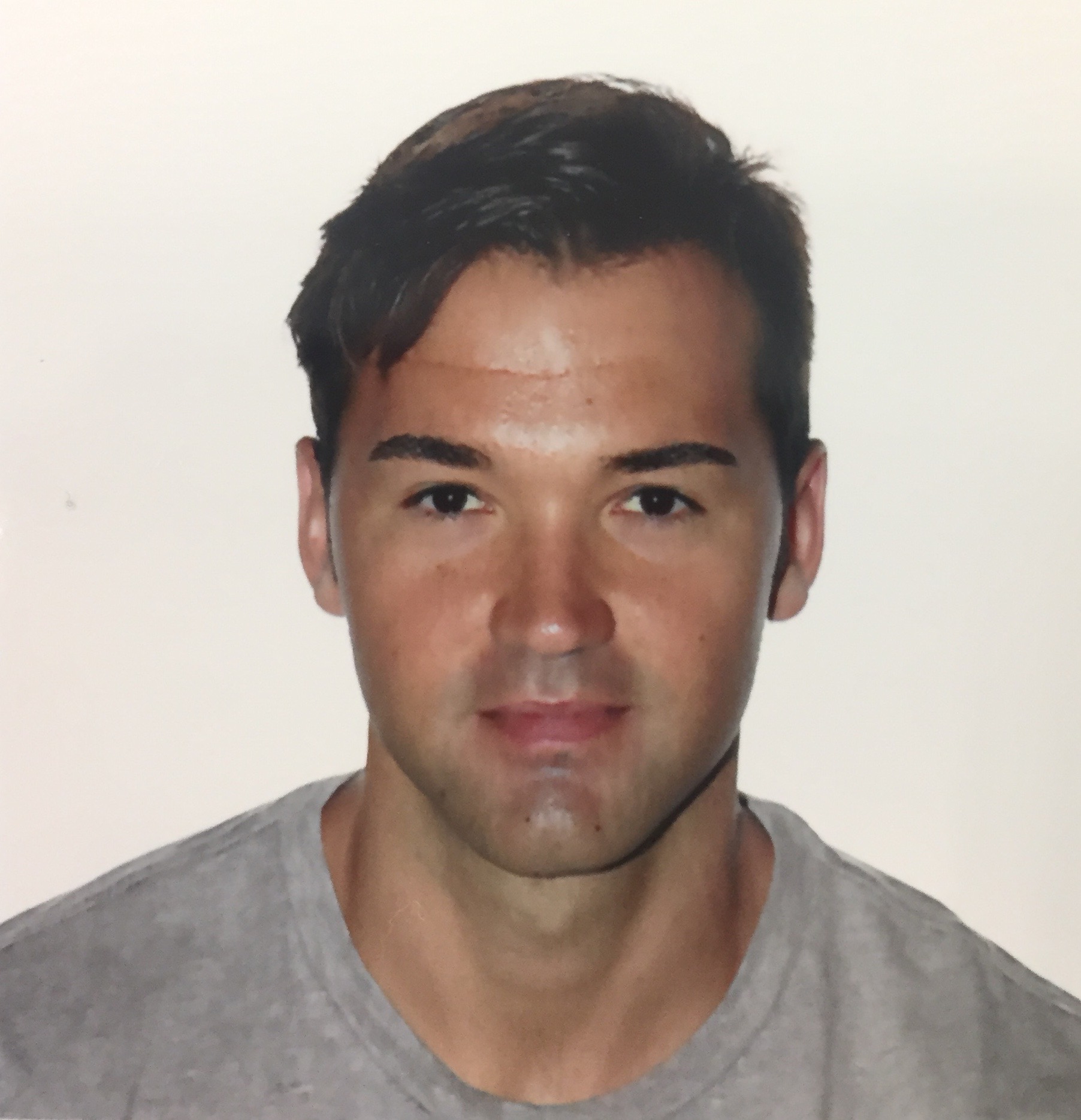}}]{Alberto Huertas Celdrán} received the MSc and PhD degrees in Computer Science from the University of Murcia, Spain. He is currently a postdoctoral fellow at the Communication Systems Group CSG, Department of Informatics IfI at the University of Zurich UZH. His scientific interests include IoT, BCI, cybersecurity, data privacy, continuous authentication, semantic technology, and computer networks.
\end{IEEEbiography}

\begin{IEEEbiography}[{\includegraphics[width=1in,clip]{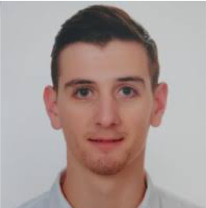}}]{Pedro M. Sánchez Sánchez} received the MSc degree in computer science from the University of Murcia, Spain. He is currently pursuing his PhD in computer science at University of Murcia. His research interests are focused on continuous authentication, networks, 5G, cybersecurity and the application of machine learning and deep learning to the previous fields.
\end{IEEEbiography}

\begin{IEEEbiography}[{\includegraphics[width=1in,clip]{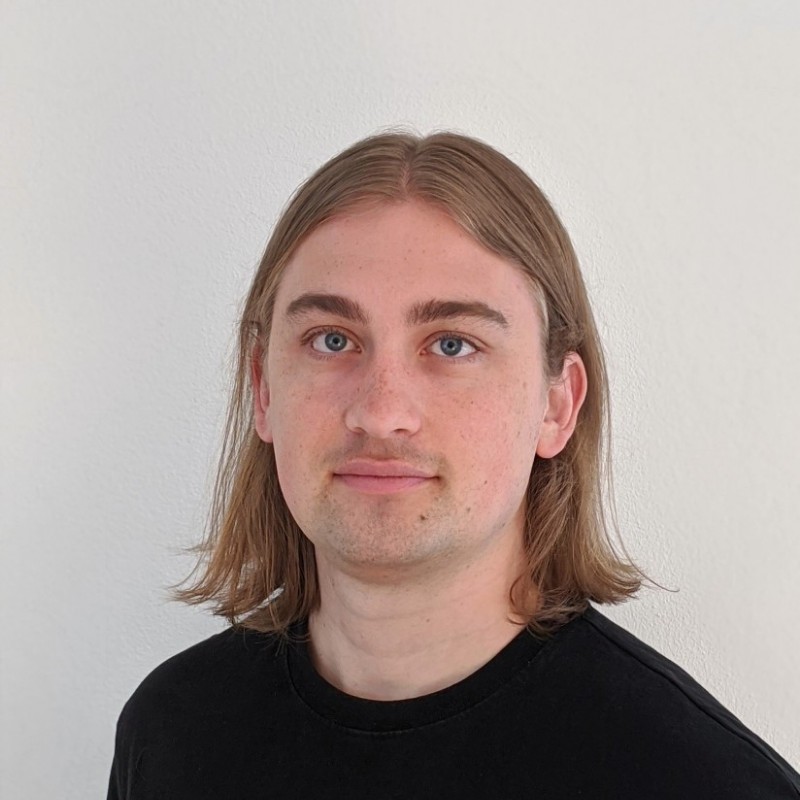}}]{Jan von der Assen} received his MSc degree in Informatics from the  University of Zurich, Switzerland. Currently, he is pursuing his Doctoral Degree under the supervision of Prof. Dr. Burkhard Stiller at the  Communication Systems Group, University of Zurich. His research interest lies at the intersection between risk management and the mitigation of cyber threats.
\end{IEEEbiography}

\begin{IEEEbiography}[{\includegraphics[width=1in,height=1.25in,clip]{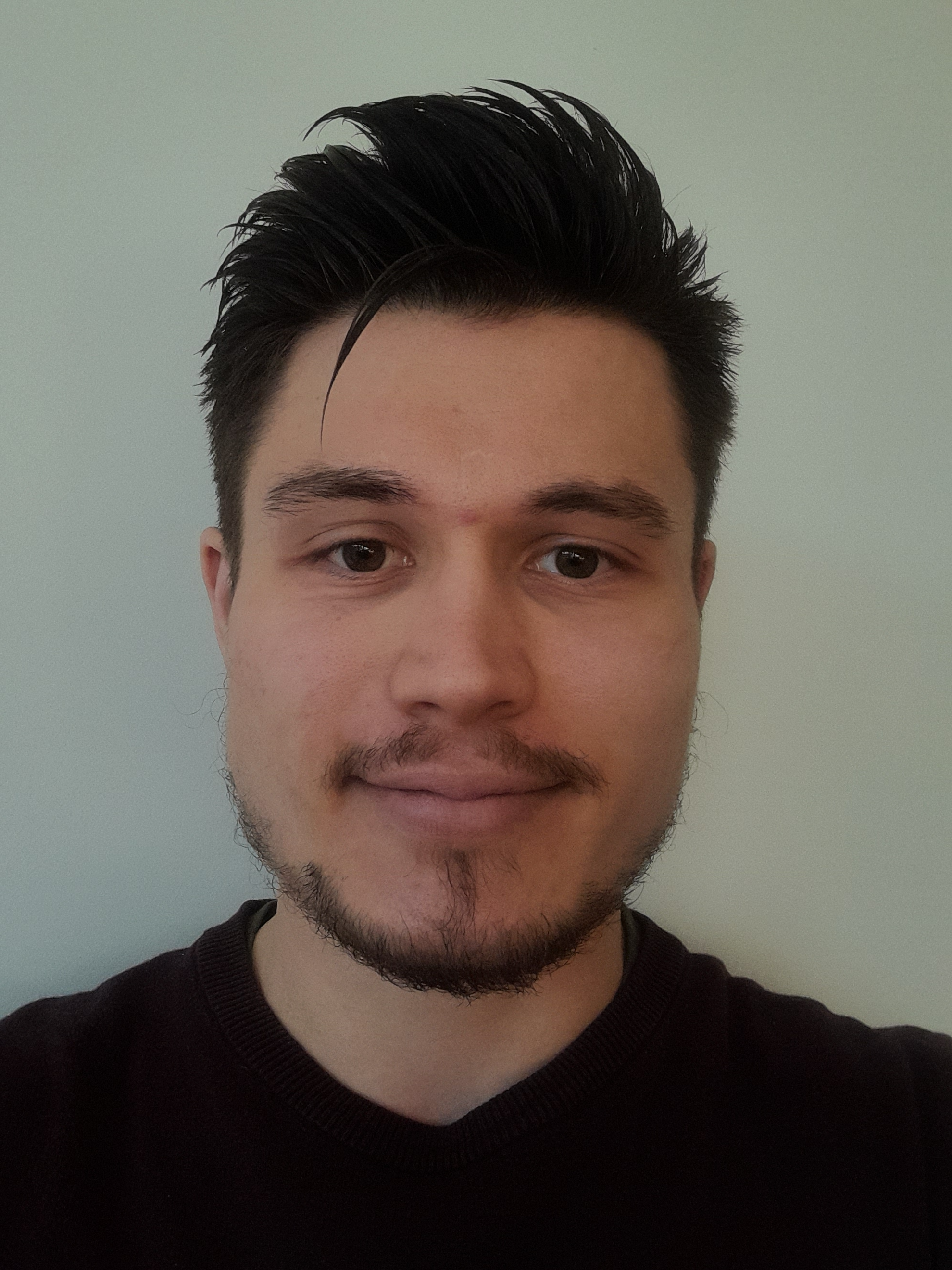}}]{Timo Schenk} received his MSc degree in Computer Science from the University of Zurich UZH, Switzerland. While having received his BSc degree from the Department of Informatics IFI UZH, he has also gained experience across multiple positions in the software engineering industry. He is passionate about cybersecurity and artificial intelligence and has a particular scientific interest in application areas where these two fields intersect.
\end{IEEEbiography}

\begin{IEEEbiography}[{\includegraphics[width=1in,clip]{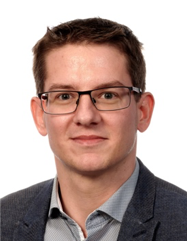}}]{Gérôme Bovet} received his Ph.D. in networks and computer systems from Telecom ParisTech, France, in 2015, and an Executive MBA from the University of Fribourg, Switzerland in 2021. He is the head of data science for the Swiss Department of Defense, where he leads a research team and portfolio of about 30 Cyber-Defence projects. His work focuses on ML and DL approaches, with an emphasis on anomaly detection, adversarial and collaborative learning applied to data gathered by IoT sensors.
\end{IEEEbiography}

\begin{IEEEbiography}[{\includegraphics[width=1in,clip]{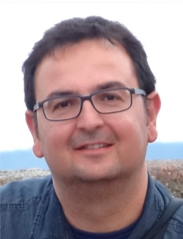}}]{Gregorio Martinez Pérez} is Full Professor in the Department of Information and Communications Engineering of the University of Murcia, Spain. His scientific activity is mainly devoted to cybersecurity and networking. He is working on different national (14 in the last decade) and European IST research projects (11 in the last decade) related to these topics, being Principal Investigator in most of them. He has published 200+ papers in international conference proceedings, magazines and journals.
\end{IEEEbiography}

\begin{IEEEbiography}[{\includegraphics[width=1in,clip]{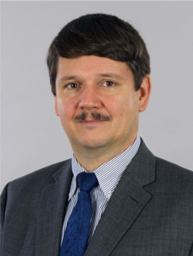}}]{Burkhard Stiller} received his MSc degree in Computer Science and the PhD degree from the University of Karlsruhe, Germany, in 1990 and 1994. Since 2004 he chairs the Communication Systems Group CSG, Department of Informatics IfI, University of Zürich UZH, Switzerland as a Full Professor. His main research interests are published in +300 papers and include decentralized systems with fully control, network and service management, IoT, and telecommunication economics.
\end{IEEEbiography}

\end{document}